\documentclass[]{spie}  %>>> use for US letter paper
\usepackage{amsmath,amsfonts,amssymb}
\usepackage{graphicx}
\usepackage[colorlinks=true, allcolors=blue]{hyperref}
\usepackage{physics}
\usepackage{setspace}
\usepackage{tocloft}
\usepackage{lineno}
\usepackage{caption} 
\usepackage{tabularray}
\usepackage{siunitx}
\usepackage{gensymb}
\usepackage{array}
\usepackage{float} 

\newcommand{\bs}[1]{\boldsymbol{#1}}

\title{The Space Coronagraph Optical Bench (SCoOB): 6. demonstration of Lyot low order wavefront control combined with high order wavefront control using a vortex coronagraph}

\author[a]{Kian Milani}
\author[b]{Kyle Van Gorkom}
\author[c]{Christopher B. Mendillo}
\author[b]{Ramya Anche}
\author[d]{Jaren N. Ashcraft}
\author[a]{Kevin Derby}
\author[b]{Jared R. Males}
\author[b]{Adam Schilperoort}
\author[b]{Ewan S. Douglas}

\affil[a]{University of Arizona, Wyant College of Optical Sciences, Tucson, Arizona, United States of America}
\affil[b]{University of Arizona, Steward Observatory, Tucson, Arizona, United States of America}
\affil[c]{University of Massachussetts Lowell, Lowell, Massachussetts, United States of America}
\affil[d]{Department of Physics, University of California, Santa Barbara, United States of America}

\authorinfo{Corresponding author: Kian Milani, kianmilani@arizona.edu}

\begin{document} 
\maketitle

\begin{abstract}
To reach and maintain high contrast levels, coronagraph instruments will require a combination of low-order and high-order wavefront control techniques to correct for dynamic wavefront error. Efficient low-order wavefront sensing and control (LOWFSC) schemes use the starlight rejected by the coronagraph such that LOWFSC can operate with a relatively bright signal to correct rapid disturbances. Meanwhile, a family of high-order wavefront sensing and control (HOWFSC) techniques utilizing the science camera have been developed to create regions of high contrast known as dark holes. These two control loops must operate simultaneously for dark holes to be maintained over long observation periods. Using a 952 actuator MEMS deformable mirror and a vector vortex coronagraph (VVC) on the Space Coronagraph Optical Bench (SCoOB), we demonstrate a Lyot-based LOWFSC loop operating in combination with a HOWFSC loop. For these experiments, implicit electric field conjugation (iEFC) is the chosen HOWFSC technique, and we demonstrate how this empirical method can be calibrated and deployed while the LOWFSC loop corrects for dynamic wavefront error. We show this combination of LOWFSC and iEFC maintained 1E-8 contrast levels in air.   
\end{abstract}

\keywords{coronagraph, dark hole, deformable mirrors, contrast}

%%%%%%%%%%%%%%%%%%%%%%%%%%%%%%%%%%%%%%%%%%%%%%%%%%%%%%%%%%%%%%%%%%%%%%%%%%%%%%%%%%%%%%%%%%%%%%%%%%%%%%%%%%%%%%%%%%%%%%%%%%%%%%%%%%%%%%%%%%%%%%%%%%%%%%%%%%%%%%%%%%%%%%%%%%%%%%%%%%%%%%%%%%%%%%%%%%%%%%%%%%%%%%%%%%%%%%%%%%%%%%%%%%%%%%%%%%%%%%%%%%%%%%%%%%
\section{Introduction}
\label{sec:llowfsc-intro}

With thousands of exoplanets having been discovered over the past few decades primarily via indirect detection methods, many astronomers have sought instruments that would be able to directly image exoplanets by creating high contrast levels in the image plane. Originally, coronagraphs were developed by Bernard Lyot to block out the light of the Sun in order to image and study coronas around our own star, but the concept of a coronagraphs has been adapted and advanced for applications in exoplanet imaging. Part of the advancement of coronagraphs has been the addition of deformable mirrors (DMs) that can be used for high-order wavefront sensing and control (HOWFSC) methods. At this point, a family of HOWFSC strategies have been developed in order to create specific regions of high contrast known as dark holes. Of these HOWFCS method,s electric field conjugation (EFC) is the most prominent and the fundamental concept is to destructively interfere the electric field of starlight that leaks into the final image plane and results in structures known as ``speckles''. 

But equally as important as HOWFSC will be a low-order wavefront sensing and control (LOWFSC) strategy capable of stabilizing low-order aberrations such that HOWFSC can achieve optimal performance. Unlike the HOWFSC algorithms, which are expected to operate at a relatively slow cadence due to the speckles being caused by quasi-static high-order aberrations\cite{males_mysterious_2021}, a low-order wavefront sensing and control (LOWFSC) method will need to operate much more quickly to correct for telescope jitter and other dynamic low-orders. 

\begin{figure}[h!]
    \centering
    \includegraphics[scale=0.3]{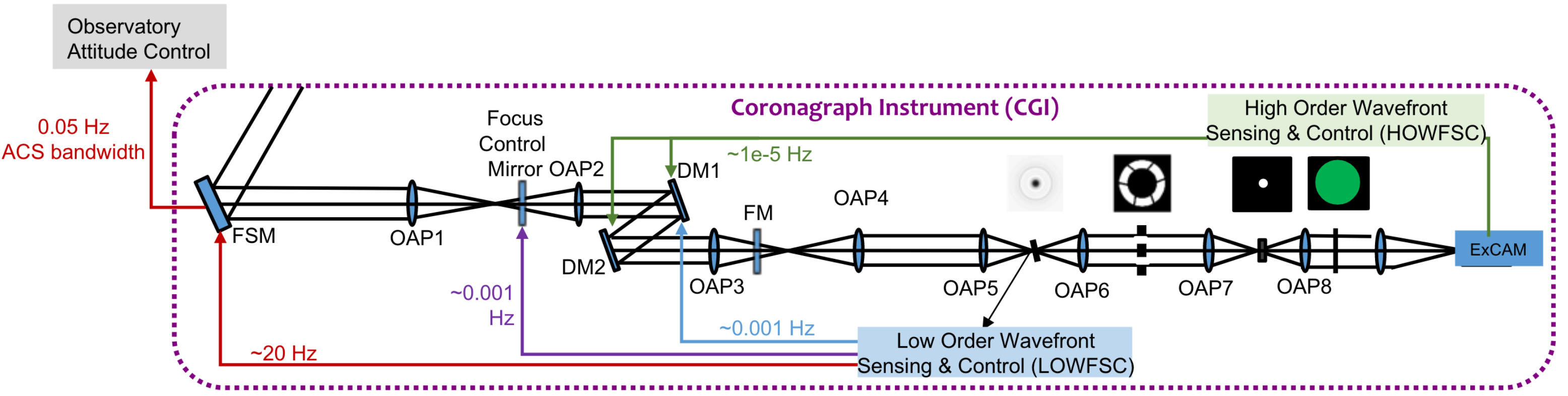}
    \caption{Unfolded layout of the Roman Coronagraph optical design from Poberezhskiy et al. 2021\cite{poberezhskiy_roman_2021} that illustrates the various control-loops, including the LOWFSC system designed for a 20Hz bandwidth for beam steering and 0.001Hz bandwidth for low order Zernikes.}
    \label{fig:llowfsc-roman-example}
\end{figure}

As an example, the Roman Coronagraph will be utilizing a LOWFSC method that operates at 1000Hz to correct line of sight (pointing) accuracy up to a bandwidth of 20Hz\cite{poberezhskiy_roman_2021}. Illustrated in Figure \ref{fig:llowfsc-roman-example}, the Roman Coronagraph uses the relatively bright signal of the on-axis star by reflecting the light that is rejected by the focal plane mask and using it to sense low-order aberrations\cite{dube_exascale_2022}. This same LOWFSC method will be utilized to stabilize the next 8 Zernike modes ranging from defocus to spherical aberration. However, the low-order Zernike corrections will be done at a much slower cadence of 0.1Hz with an operational bandwidth of 0.0016Hz (about 1 cycle per 10 min). This choice was made because STOP modeling for the Roman Coronagraph predicted much slower drifts of those Zernike terms\cite{seo_demonstration_2025}. 

For a coronagraph instrument to operate at high contrast levels, some form of LOWFSC and HOWFSC need to be combined such that a dark hole can be created and maintained for extended observation periods. While the Roman Coronagraph has developed and tested a strategy for an HLC and SPC, the same approach cannot be applied to all coronagraphs due to hardware differences of the FPMs. Specifically, the FPMs of the Roman Coronagraph are designed with a central "dimple" that acts as a Zernike WFS\cite{dube_exascale_2022}. The on-axis signal rejected by the FPM is reflected into the LOWFSC Optical Barrel Element (LOBE) where it is recollimated to produce a relayed pupil image at the LOWFSC camera. Through simulations and testing, this approach was demonstrated to be successful for the Roman coronagraph requirements and architecture\cite{seo_demonstration_2025}, but this same approach cannot be applied to all coronagraph types. 

One example is the vortex coronagraph where the FPM is a phase mask that may not have central obscuration. This means the on-axis signal is not  directly rejected by the FPM, but by the Lyot stop. So using a reflective Lyot stop (RLS), this rejected light can still be used to perform Lyot-LOWFSC (LLOWFSC). This approach has been demonstrated in works such as Singh et al. 2015\cite{singh_-sky_2015} and Mendillo et al. 2023\cite{mendillo_reflective_2023}, but has yet to be combined with a HOWFSC method demonstrating low-order stabilization with dark hole creation simultaneously. 

Here, we test this LLOWFSC approach using the vector vortex coronagraph (VVC) on the Space Coronagraph Optical Bench (SCoOB) and simultaneously use implicit EFC (iEFC) to create and maintain a dark hole. Section \ref{sec:llowfsc-theory} details the mathematical theory for LLOWFSC and Section \ref{sec:llowfsc-combined-sims} tests this approach using simulations with the SCoOB hardware parameters. Lastly, Section \ref{sec:llowfsc-combined-scoob} presents the results after deploying LLOWFSC on SCoOB to correct dynamic WFE that is injected with the BMC Kilo-DM. Simultaneously, iEFC is deployed and a contrast of $9.6\times10^{-10}$ is maintained over the course of a 2 hr period of data collection. Critically, we demonstrate that contrast would be limited to $>10^{-7}$ if LLOWFSC were not used. Similarly, contrast would be limited to $5\times10^{-8}$ if LLOWFSC and HOWFSC are not combined with the reference offset approach discussed below. 

%%%%%%%%%%%%%%%%%%%%%%%%%%%%%%%%%%%%%%%%%%%%%%%%%%%%%%%%%%%%%%%%%%%%%%%%%%%%%%%%%%%%%%%%%%%%%%%%%%%%%%%%%%%%%%%%%%%%%%%%%%%%%%%%%%%%%%%%%%%%%%%%%%%%%%%%%%%%%%%%%%%%%%%%%%%%%%%%%%%%%%%%%%%%%%%%%%%%%%%%%%%%%%%%%%%%%%%%%%%%%%%%%%%%%%%%%%%%%%%%%%%%%%%%%%
\section{The Lyot-based LOWFSC Method}
\label{sec:llowfsc-theory}

Discussed in previous works from Ashcraft et al.\cite{ashcraft_space_2022} and Van Gorkom et al.\cite{van_gorkom_space_2024}, SCoOB is a coronagraphic testbed developed at the University of Arizona to advance technologies needed for high-contrast imaging. This includes coronagraphic masks, detectors, and wavefront control methods. For this work, SCoOB uses a VVC that imparts the vortex phase profile using liquid crystal polymers acting as a half-wave plate with a spatially varying optic axis\cite{mawet_annular_2005}. The beam after the vortex FPM is relayed to the RLS where the geometric pupil diameter is 9.1 mm. This RLS is transmissive in the central region inside the geometric pupil, but reflective outside the pupil to direct the rejected on-axis light into the LLOWFSC branch of SCoOB. Here, a lens is used to focus the vortex beam onto a detector (CAMLO) in order to collect more photons per pixel and achieve better signal to noise. For this setup, the nominal focal length of the lens is 200 mm, but CAMLO is slightly defocused to improve the linear regime of LLOWFSC while remaining in the desired camera region of interest (ROI).

\begin{figure}[h!]
    \centering
    \includegraphics[scale=0.325]{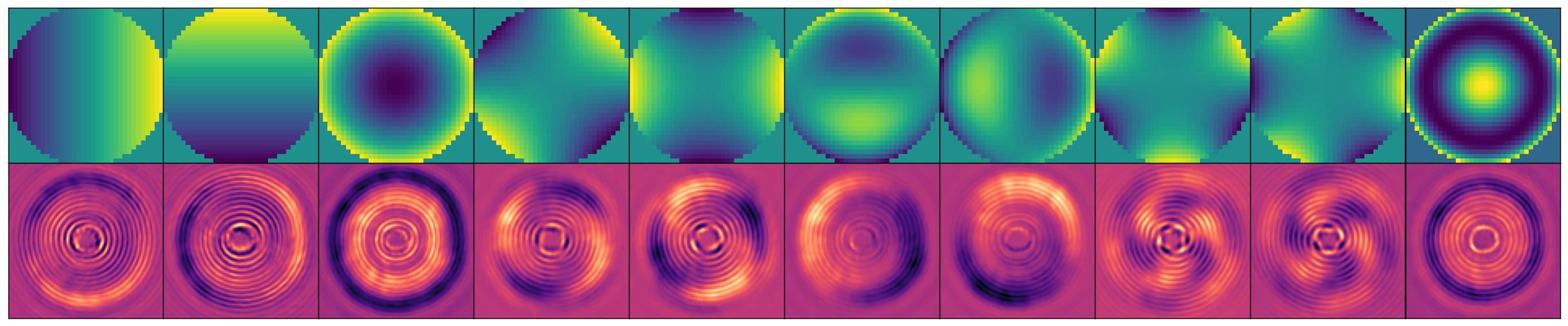}
    \caption{Examples of simulated LLOWFSC responses to each of the first 10 Zernike modes.}
    \label{fig:sim-llowfsc-responses}
\end{figure}

The concept of this LLOWFSC approach is to stabilize low-order aberrations by computing and applying DM commands that drive the CAMLO image to a known ``good'' reference image ($I_{ref}$). In order to compute the DM commands for each LLOWFSC iteration, we empirically measure the response of images on CAMLO to a set of low-order Zernike modes. Examples of these responses are illustrated in Figure \ref{fig:sim-llowfsc-responses}. Following equations \ref{eq:llowfsc-response}-\ref{eq:llowfsc-control-matrix}, each Zernike response ($\mathbf{R_{Z,i}}$) is measured with a central difference approximation after applying the mode to the DM. The responses are then stored into the Zernike response matrix $G_{LO}$ and a pseudo-inverse is used to compute the control matrix $P_{LO}$. For the work presented here, we only consider the first 10 Zernike modes (up to spherical aberration) for LLOWFSC. 

\begin{equation}
    \boldsymbol{R_{Z,i}} = \frac{\boldsymbol{I_{Z,i,p}} - \boldsymbol{I_{Z,i,n}}}{2a_{calib}}
    \label{eq:llowfsc-response}
\end{equation}

\begin{equation}
    G_{LO} = 
    \begin{bmatrix}
        \mathbf{R_{Z,1}} & 
        \mathbf{R_{Z,2}} & 
        \cdots & 
        \mathbf{R_{Z,i}} & 
        \cdots & 
        \mathbf{R_{Z,N_{modes}}}
    \end{bmatrix}
    \label{eq:llowfsc-response-matrix}
\end{equation}

\begin{equation}
    P_{LO} = (G_{LO}^\top G_{LO})^{-1} G_{LO}^\top
    \label{eq:llowfsc-control-matrix}
\end{equation}

During each LLOWFSC iteration, the DM commands are computed by measuring the difference between the acquired CAMLO image and previously measured reference image. 

\begin{equation}
    \delta I = I_{CAMLO} - I_{ref}
    \label{eq:delta-im}
\end{equation}

The difference image is concatenated into a column vector ($\bs{\delta I}$) so that we can multiply by the control matrix, which gives us the modal coefficients for each Zernike mode that was calibrated. The update to the DM actuators ($\delta A$) is computed from these coefficients after applying modal gain coefficients ($\boldsymbol{g}$) to improve the control-loop stability. In some scenarios, an additional leakage term ($l$) can also be used to remove built up DM commands, but later results set this leakage to 0. 

\begin{equation}
    \boldsymbol{\delta C_{LO}} = \boldsymbol{g} \cdot P_{LO} \boldsymbol{\delta I}
    \label{eq:modal-coeff}
\end{equation}

\begin{equation}
    \delta A = \sum_{i=1}^{N_{modes}} \boldsymbol{\delta C_{LO,i}} A_{Z,i}
    \label{eq:llowfsc-udpate}
\end{equation}

\begin{equation}
    A = (1 - l) A_{old} + \delta A
    \label{eq:llowfsc-command}
\end{equation}

Because LLOWFSC is calibrated before creating a dark hole, the challenge to overcome when combining LLOWFSC with a HOWFSC method such as iEFC is to prevent LLOWFSC from mistakenly removing components of the HOWFSC command that couple into the LLOWFSC signal. To do so, a separate process is introduced to predict the reference offset ($\delta I_{ref}$) that any HOWFSC command would induce in the LLOWFSC signal. So when the reference offset is taken into account, the total LLOWFSC difference measurement becomes

\begin{equation}
    \delta I = I_{CAMLO} - (I_{ref} + \delta I_{ref})\text{.}
    \label{eq:offset-delta-im}
\end{equation}

\noindent How the reference offset can be predicted is discussed in the following sections; first with simulations to nominally test the methodology and then with a complete experiment on SCoOB demonstrating the combination of LLOWFSC with iEFC. 

%%%%%%%%%%%%%%%%%%%%%%%%%%%%%%%%%%%%%%%%%%%%%%%%%%%%%%%%%%%%%%%%%%%%%%%%%%%%%%%%%%%%%%%%%%%%%%%%%%%%%%%%%%%%%%%%%%%%%%%%%%%%%%%%%%%%%%%%%%%%%%%%%%%%%%%%%%%%%%%%%%%%%%%%%%%%%%%%%%%%%%%%%%%%%%%%%%%%%%%%%%%%%%%%%%%%%%%%%%%%%%%%%%%%%%%%%%%%%%%%%%%%%%%%%%
\section{Combined LLOWFSC and iEFC Simulations}
\label{sec:llowfsc-combined-sims}

\begin{figure}[h!]
    \centering
    \includegraphics[scale=0.325]{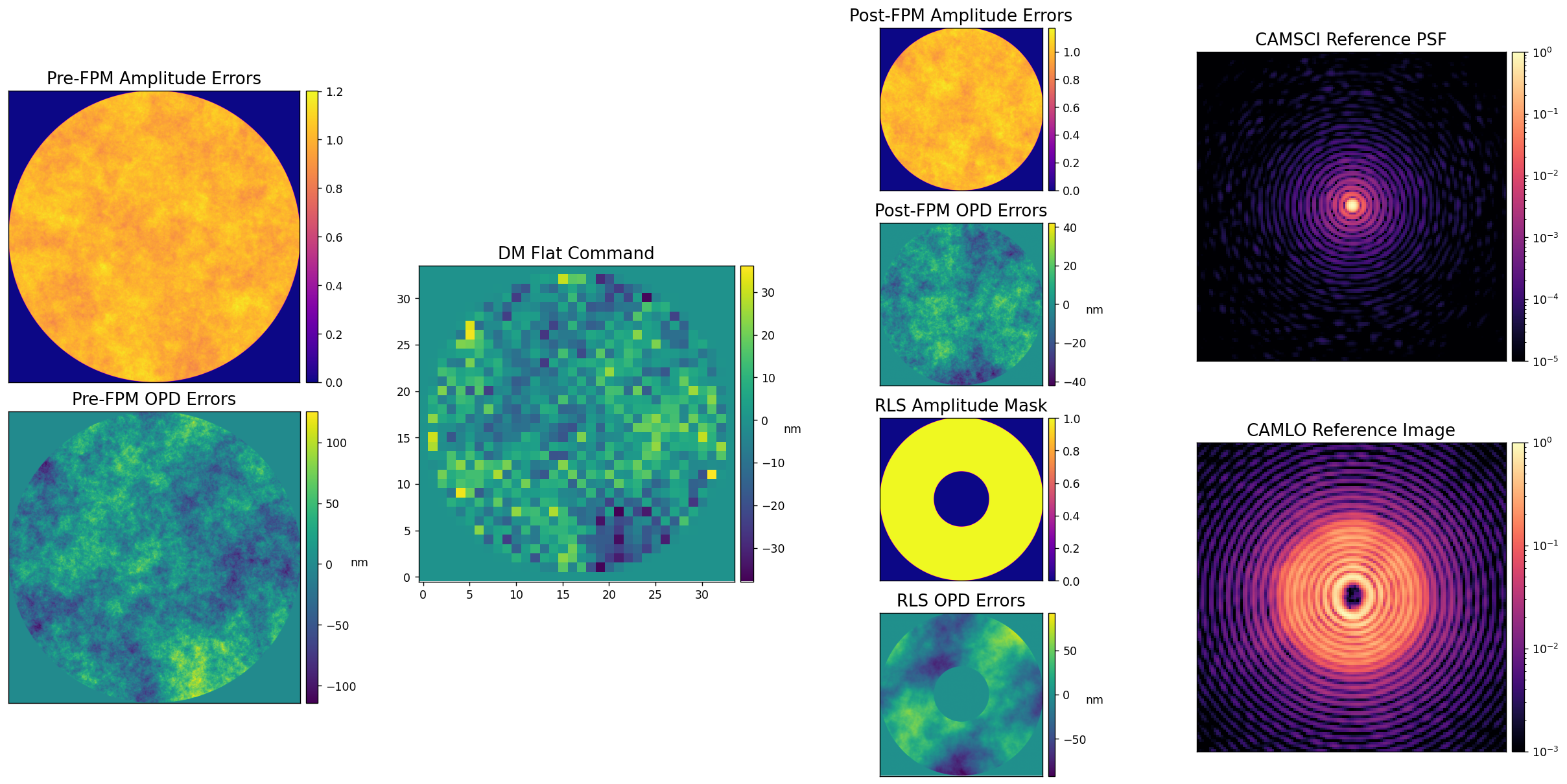}
    \caption{Setup of the Fraunhofer model for SCoOB that simulates both CAMSCI images and CAMLO images using two branches in the model. Each branch shares a pre-FPM WFE shown to the left along with the DM flat command that simulates the use of phase retrieval to correct for the pre-FPM OPD. But after propagating through the vortex FPM, the CAMSCI branch uses an additional post-FPM WFE to simulate the backend aberrations while the CAMLO branch uses a separate WFE defined specifically for the RLS path of SCoOB.}
    \label{fig:sim-errors}
\end{figure}

We first use a Fraunhofer model of the SCoOB instrument to test this method of combining LLOWFSC with a HOWFSC method. Illustrated in Figure \ref{fig:sim-errors}, the model includes two branches, one used to simulate science camera (CAMSCI) images and another to simulate LLOWFSC camera (CAMLO) images. Included in both branches is the pre-FPM wavefront error along with a DM flat command that simulates the use of phase retrieval to flatten the OPD before the FPM. The model propagates the pupil plane wavefront through the vortex phase mask using the same method presented in Krist et al. 2019\cite{krist_numerical_2019}. 

After propagating through the vortex, the CAMSCI branch includes the Lyot stop aperture and additional wavefront error that is caused by optics downstream of the FPM. A final MFT is used to simulate CAMSCI images at the pixelscale of the SCoOB hardware, which is about 0.307$\lambda/D$. 

In the CAMLO branch, the inverse of the Lyot stop aperture is used to simulate the RLS, but restricted to an outer diameter of 1 in. Before the RLS transmission is applied to the wavefront, we perform a back propagation to the plane of the recollimating OAP after the vortex using angular spectrum propagation. This back propagation is done because the OAP has a smaller diameter of 15 mm, so we back propagate to apply the OAPs aperture outside the pupil plane, then propagate back to the pupil where the transmission and OPD of the RLS is applied to the wavefront. At this point, the transfer function of the camera's defocus is applied to the wavefront and a final MFT is used to simulate the CAMLO images at a pixelscale of 0.270$\lambda/D$. 

\begin{figure}[h!]
    \centering
    \includegraphics[scale=0.5]{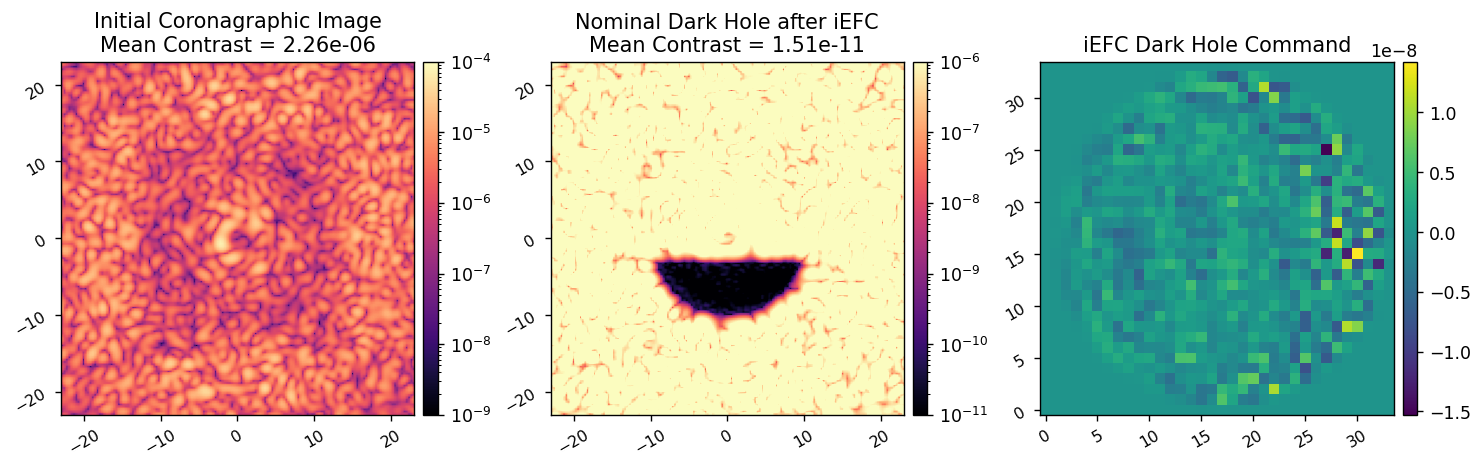}
    \caption{To the left is the initial coronagraphic image from the CAMSCI branch of the SCoOB model with only the DM flat command applied. In the middle is the final dark hole image produced after running iEFC with the final dark hole command shown to the right. Note the total DM command is a sum of the DM flat and this dark hole command.}
    \label{fig:sim-iefc}
\end{figure}

Using this SCoOB model, we perform iEFC to create a nominal dark hole. For this iEFC simulation, we use a basis of Hadamard modes as those are the most frequent choice for in lab experiments \cite{van_gorkom_space_2024} and are used for the SCoOB tests in Section \ref{sec:llowfsc-combined-scoob}. Figure \ref{fig:sim-iefc} illustrates the initial coronagraphic image starting with just the DM flat command followed by the final dark hole and the command that produced it. 

\begin{figure}[h!]
    \centering
    \includegraphics[scale=0.35]{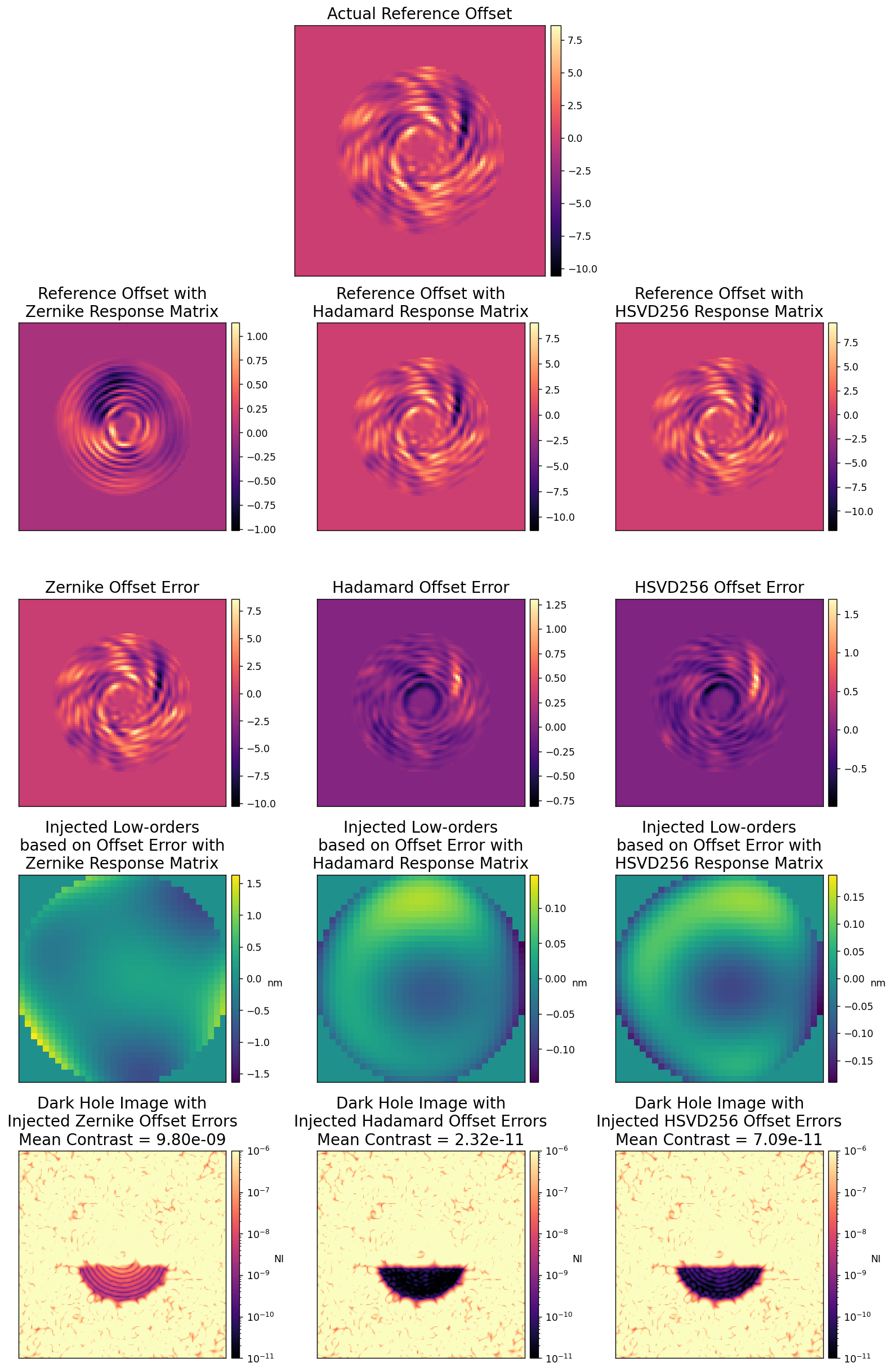}
    \caption{At the top is the ``true'' reference offset that should be used in order to prevent the iEFC command from coupling into LLOWFSC. In the second row is the predicted reference offset with the Zernike, Hadamard, and HSVD256 response matrices. In the third row is the relative error of the predicted offsets. In the fourth row is the residual DM command each offset error would induce. Finally, row 5 contains final dark hole result with the residual DM command added. This illustrates that the Zernike response matrix would not be adequate for predicting the reference offset, but the Hadamard and HSVD256 basis would be sufficient.}
    \label{fig:sim-ref-offsets}
\end{figure}

The final dark hole command from the iEFC simulation is then used to test the reference offset methodology by evaluating the accuracy of the reference offset as presented in Figure \ref{fig:sim-ref-offsets}. First, we evaluate what the ``true'' reference offset should be in the simulation. Then, we compute the predicted reference offset by multiplying the LLOWFSC response matrix by the dark hole command as shown in Eq \ref{eq:ref-offset-zer}. 

\begin{equation}
    \boldsymbol{\delta I_{ref}} = G_{LO} \boldsymbol{{A}_{DH}}
    \label{eq:ref-offset-zer}
\end{equation}

To evaluate the impact the reference offset would have on the dark hole, we compute the difference between the true offset and the predicted offset and multiply this difference by the LLOWFSC control matrix. This yields the Zernike coefficients of the DM command that LLOWFSC would not compensate for, meaning the impact of these errors would leak through to the final dark hole. Using the nominal LLOWFSC response matrix which calibrated the first 10 Zernike modes, we find that the reference offset would not be accurate enough as the final dark hole contrast is degraded from $1.5\times10^{-11}$ to $9.8\times10^{-9}$. 

To solve this, a different LLOWFSC response matrix can be measured using a basis set that would more accurately capture the offset of the dark hole command. Given we use Hadamard modes as the basis for iEFC, the obvious choice is to measure a LLOWFSC response matrix with Hadamard modes as well ($G_{HAD}$). Repeating the same steps, the  reference offset with $G_{HAD}$ is computed using Eq \ref{eq:ref-offset-had}, we find this predicts a much more accurate offset as the final dark hole only degrades to a contrast of $2.3\times10^{-11}$. 

\begin{equation}
    \boldsymbol{\delta I_{ref}} = G_{HAD} \boldsymbol{{A}_{DH}}
    \label{eq:ref-offset-had}
\end{equation}

\begin{figure}[h!]
    \centering
    \includegraphics[scale=0.325]{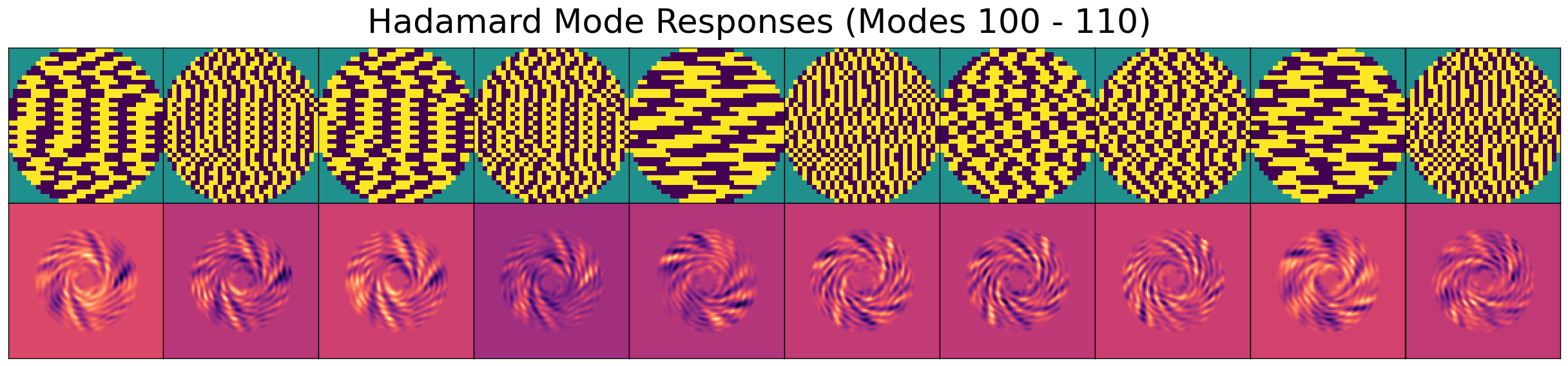}
    \caption{LLOWFSC responses for a subset of 10 Hadamard modes using the SCoOB Fraunhofer model.}
    \label{fig:sim-hadamard-responses}
\end{figure}

\begin{figure}[h!]
    \centering
    \includegraphics[scale=0.325]{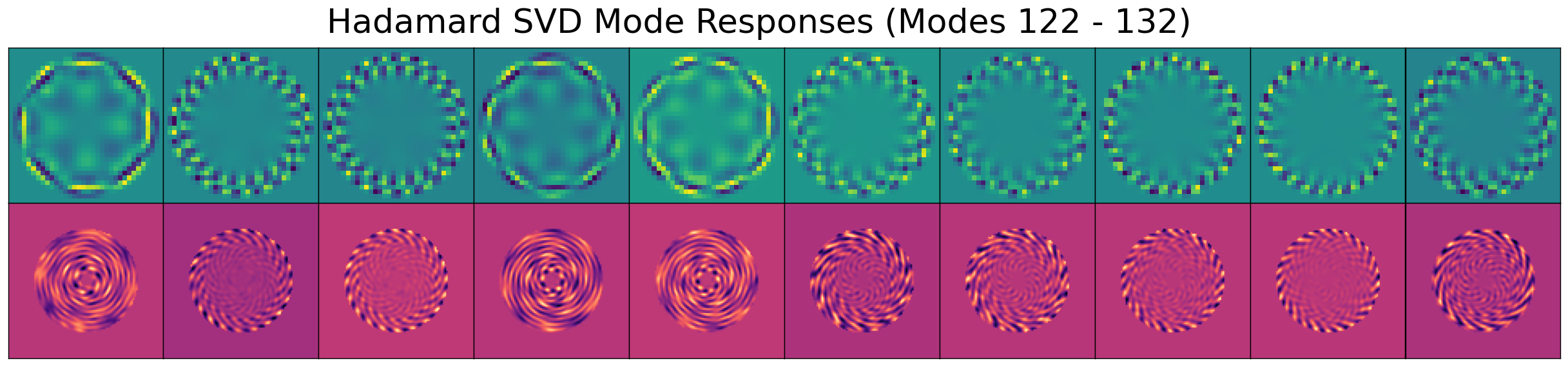}
    \caption{LLOWFSC responses for a subset of 10 modes from the Hadamard SVD using the SCoOB Fraunhofer model. Because SVD orders the modes from most sensitive to least sensitive, the Hadamard SVD basis is truncated to only 256 modes to test if this would be adequate for the reference offset needed.}
    \label{fig:sim-hsvd-responses}
\end{figure}

However, calibrating LLOWFSC with all Hadamard modes can be a disadvantage because the large number of modes will make the calibration procedure require significantly more time. A possible solution would be to perform the complete Hadamard calibration once, but then perform a singular value decomposition (SVD) with the Hadamard response matrix to select the modes most significant to the LLOWFSC signal. Here, we truncate from the original 1024 Hadamard modes to only the first 256 Hadamard SVD (HSVD) modes and generate a response matrix just for this subset ($G_{HSVD}$). Figure \ref{fig:sim-hsvd-responses} illustrates a subset of these HSVD modes along with the response measured with the SCoOB model. Following the same logic as before, the predicted reference offset is computed using this response matrix and the error from the ``true'' reference offset is evaluated with the LLOWFSC control matrix. Here, the offset error with the HSVD modes degrades the contrast to a value of $7.1\times10^{-11}$. While worse than the leakage using the complete Hadamard basis, there is a trade-off between the accuracy of the predicted reference offset and the number of modes to calibrate that must be evaluated for the contrast goals of individual instruments. These simulation results are summarized by Figure \ref{fig:sim-ref-offsets} where the reference offsets, relative offset errors, and contrast results are presented for each of the modal basis discussed. 

\begin{equation}
    \boldsymbol{\delta I_{ref}} = G_{HSVD} \boldsymbol{{A}_{DH}}
\end{equation}

\begin{figure}[h!]
    \centering
    \includegraphics[scale=0.4]{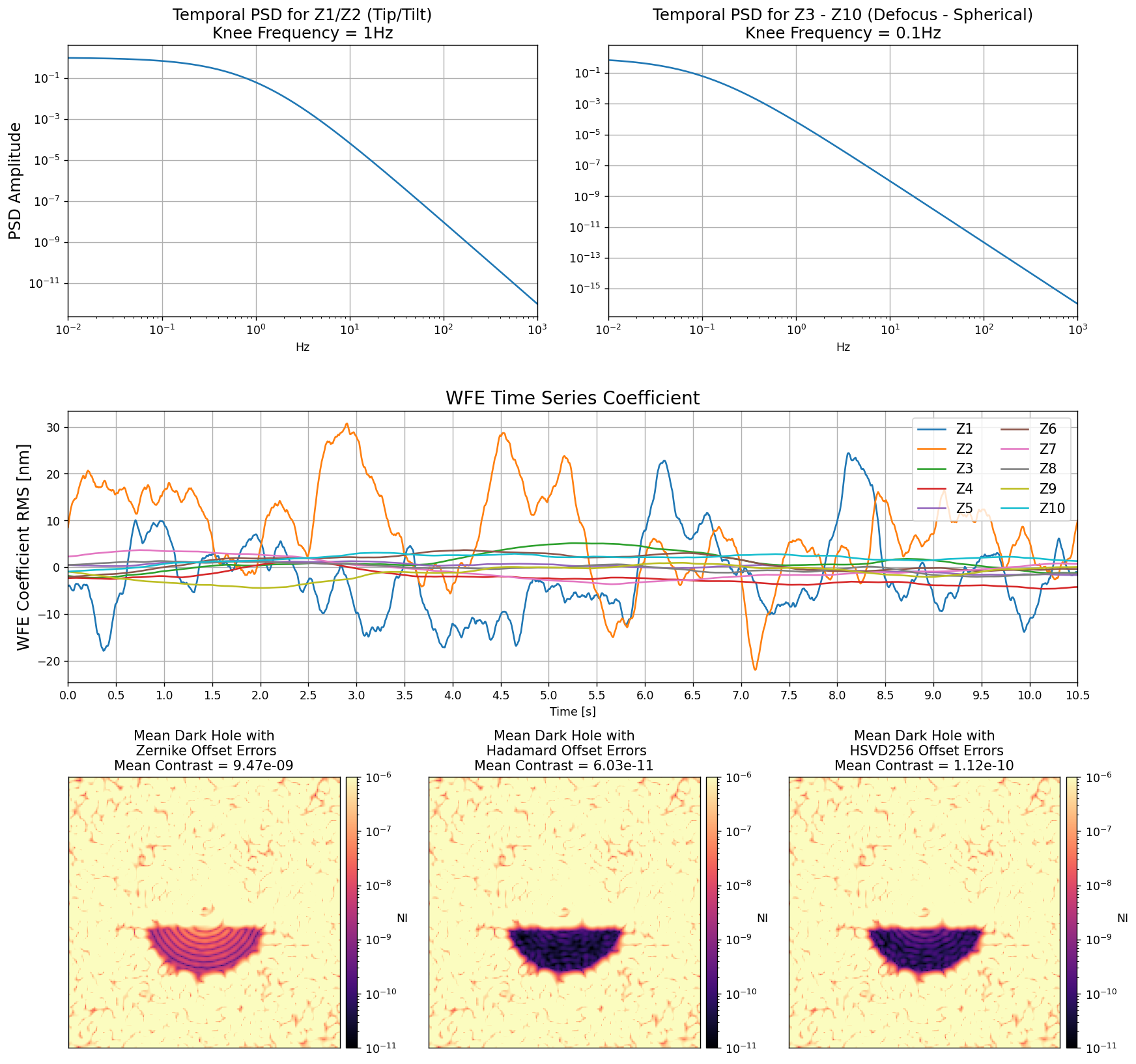}
    \caption{In the top row are the two PSDs defined for the tip/tilt and the defocus - spherical Zernike modes. In the middle is the RMS coefficient for each of the modes across the time series used for the simulation. In the final row are the average dark holes across the time series demonstrating how the reference offset with the Zernike, Hadamard, and HSVD modes will impact the achieved contrast.}
    \label{fig:sim-time-series}
\end{figure}

Lastly a time series of WFE is generated to test the LLOWFSC method with each of the reference offsets. The time series is generated by defining a power spectral density (PSD) for each of the WFE modes that will be injected in the simulation. Again, we only consider the first 10 Zernike modes so a PSD for each of these is defined with a chosen knee frequency and slope. We expect jitter to be the most dynamic and the most dominant term that LLOWFSC would need to correct, so the knee frequency for the tip/tilt modes is chosen to be 1 Hz and the RMS coefficient for the time series is 10 nm for these modes. For defocus to spherical aberration, the knee frequency is chosen to be 0.1 Hz and the RMS coefficient for each mode is chosen to be 2.5 nm. The slope of the PSD for all Zernike modes is chosen to be -4. Figure \ref{fig:sim-time-series} illustrates these two PSDs along with the time series of coefficients that are generated for each mode. For this simulation, we assume LLOWFSC is running at 500Hz, so each time series is downsampled such that the time interval between iterations is 2 ms. This automatically filters out the higher frequency content within the time series, but the intention of this simulation is to validate the reference offset methodology, not to be a fully realistic simulation.

Within each simulation iteration, a CAMLO frame is computed and used to calculate the DM correction. After the correction is applied to the simulated DM, the injected wavefront error is progressed to the next iteration to simulate a time lag of 2 ms. Lastly, a CAMSCI frame is computed to obtain the state of the dark hole at each LLOWFSC iteration. At the end of the simulation, all CAMSCI frames are averaged to evaluate the mean dark hole over the course of the time series. Doing so, we find the mean contrast with the Zernike, Hadamard and HSVD offsets to be $9.47\times10^{-9}$, $6.03\times10^{-11}$, and $1.12\times10^{-10}$ respectively. Each mean CAMSCI image is included in Figure \ref{fig:sim-time-series}.

Due to the computational complexity of simulating the vortex coronagraph, rigorously simulating the combination of LLOWFSC and iEFC during the iEFC calibration or the closed-loop stages is not feasible with available computational hardware. But, these simulations demonstrate that $10^{-10}$ contrasts can be maintained with the reference offset methodology if a separate response matrix is used to capture the high-order structure of dark hole commands. Section \ref{sec:llowfsc-combined-scoob} demonstrates a complete combination of LLOWFSC and iEFC with the SCoOB hardware. 

%%%%%%%%%%%%%%%%%%%%%%%%%%%%%%%%%%%%%%%%%%%%%%%%%%%%%%%%%%%%%%%%%%%%%%%%%%%%%%%%%%%%%%%%%%%%%%%%%%%%%%%%%%%%%%%%%%%%%%%%%%%%%%%%%%%%%%%%%%%%%%%%%%%%%%%%%%%%%%%%%%%%%%%%%%%%%%%%%%%%%%%%%%%%%%%%%%%%%%%%%%%%%%%%%%%%%%%%%%%%%%%%%%%%%%%%%%%%%%%%%%%%%%%%%%
\section{Combined LLOWFSC and iEFC SCoOB Experiments}
\label{sec:llowfsc-combined-scoob}

\begin{figure}[h!]
    \centering
    \raisebox{-0.5\height}{\includegraphics[scale=0.285]{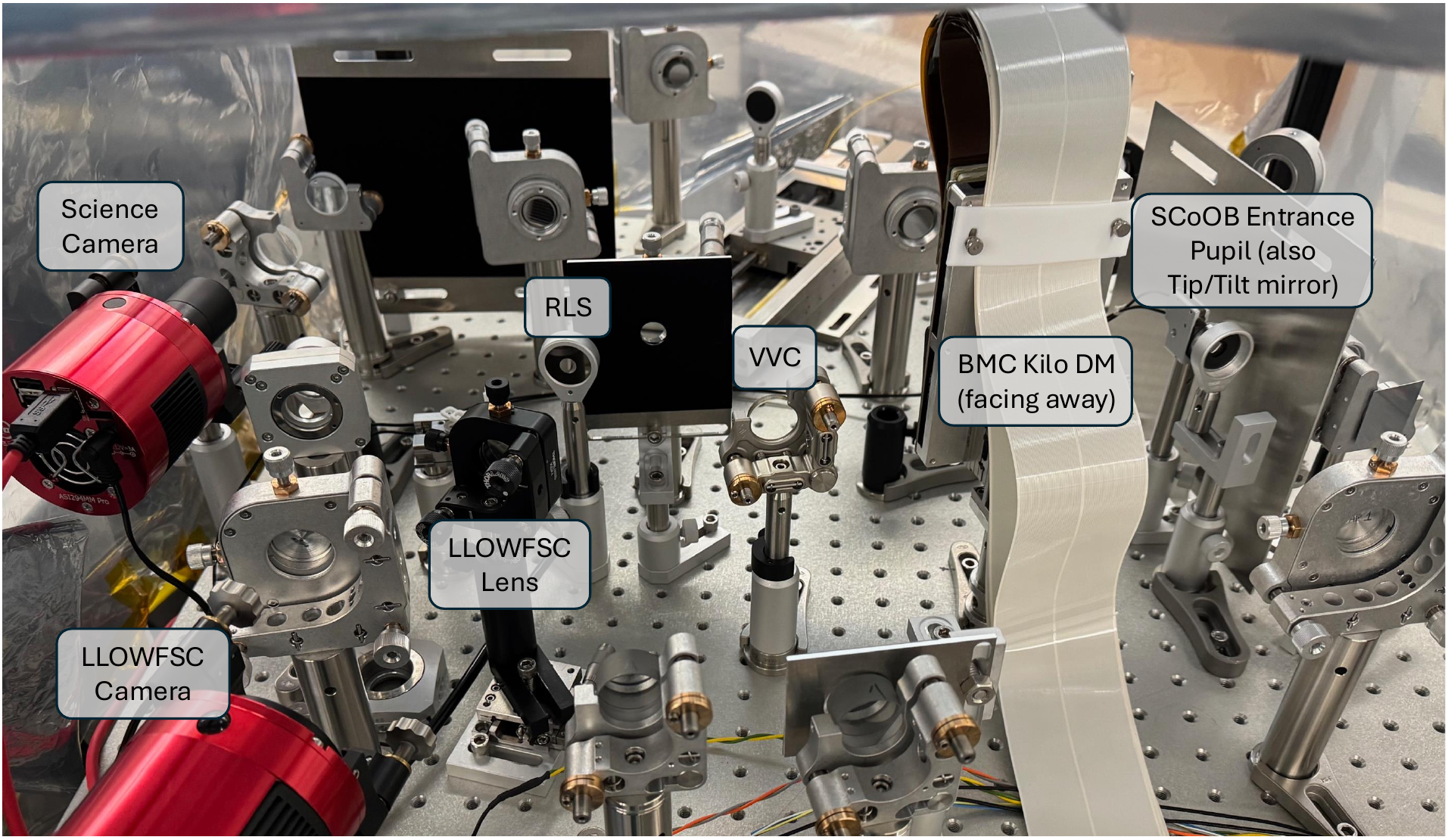}}
    \raisebox{-0.525\height}{\includegraphics[scale=0.575]{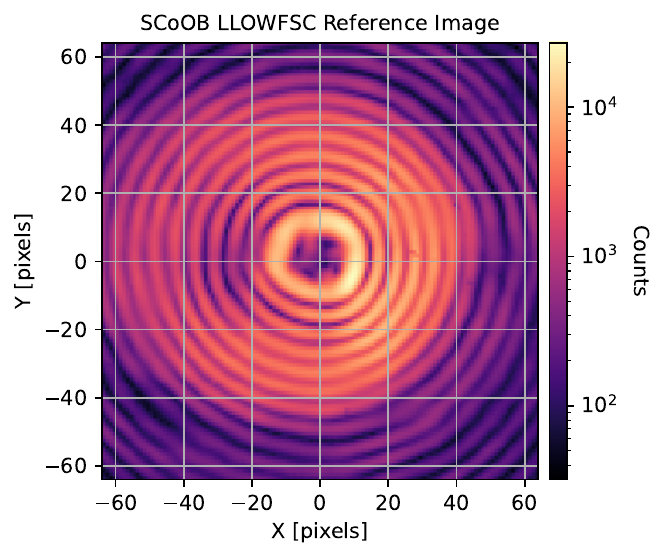}}
    \caption{On the left is the SCoOB hardware depicting where the RLS, LLOWFSC lens, and CAMLO are located relative to the other optics. On the right is the reference image recorded with CAMLO prior to creating a dark hole.}
    \label{fig:scoob-llowfsc-setup}
\end{figure}

To perform LLOWFSC on SCoOB, an RLS manufactured at University of Massachusetts Lowell is used. This RLS is a glass substrate with an annular reflective area created with deposited aluminum. Figure \ref{fig:scoob-llowfsc-setup} illustrates where this RLS is within SCoOB's optical system. It is mounted at about a 10\degree an angle with respect to the incoming beam to redirect the diffracted light to CAMLO, but the angle was not precisely measured due to frequent hardware changes made on SCoOB. Also in Figure \ref{fig:scoob-llowfsc-setup} is an example of the LLOWFSC reference image measured prior to closing the LLOWFSC control-loop. 

\begin{figure}[h!]
    \centering
    \includegraphics[scale=0.55]{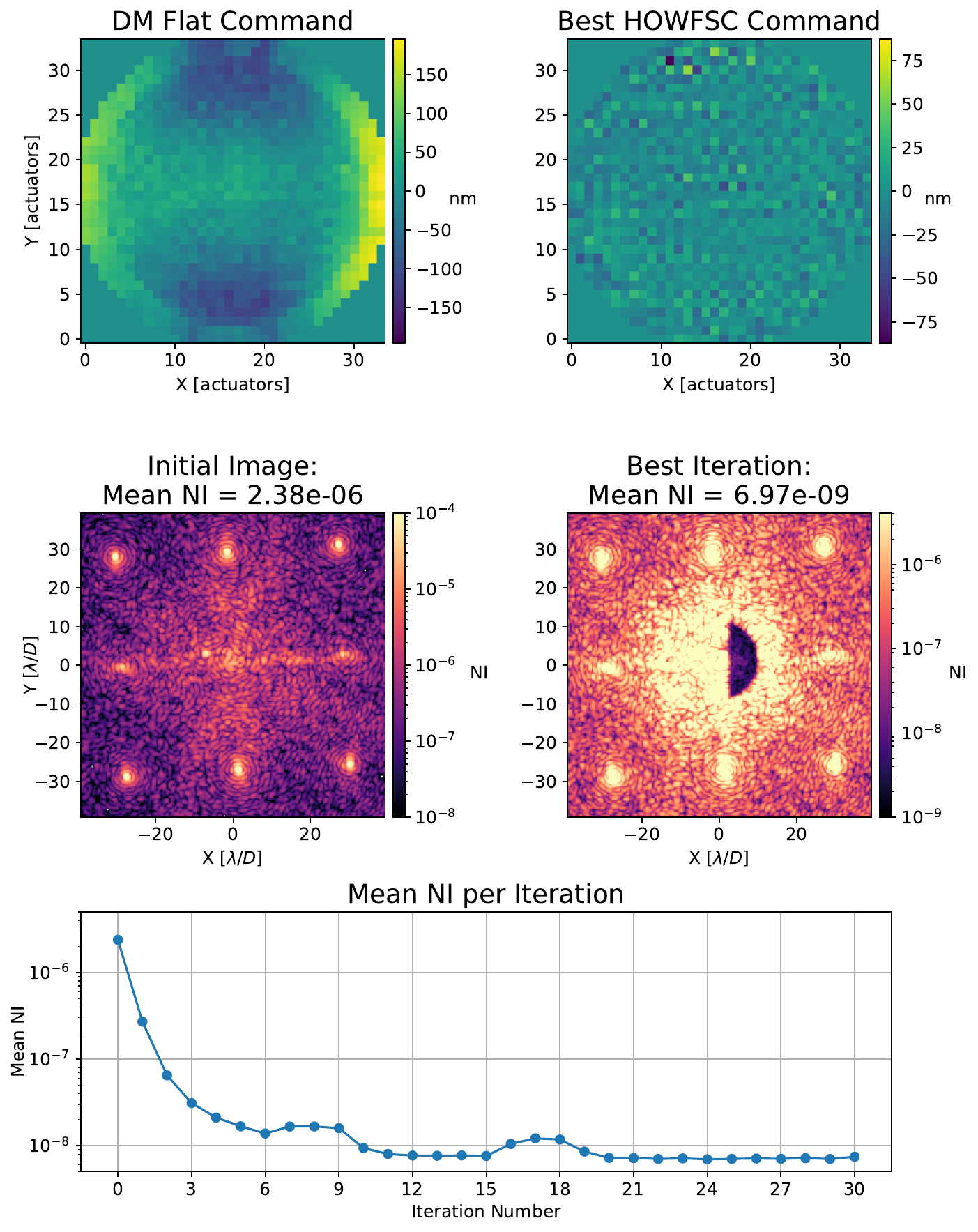}
    \caption{In the top left is the DM flat command used on SCoOB for the hardware configuration of these experiments followed by the final dark hole command generated after running iEFC. In the middle left is the initial coronagraphic image before iEFC while the middle right is the final dark hole image produced by the iEFC command. In the bottom are the contrast values at each iteration of this baseline iEFC experiment. }
    \label{fig:scoob-iefc-baseline}
\end{figure}

Prior to combining LLOWFSC and iEFC, we perform a baseline iEFC test to understand the performance of SCoOB with the current state of hardware. This iEFC experiment is conducted using two Fourier probes and a basis of 1024 Hadamard modes similar to the iEFC experiments conducted in Van Gorkom et al.\cite{van_gorkom_space_2024} Calibrating iEFC with this choice of probes and modal basis required about 1500s (25 min). Presented in Figure \ref{fig:scoob-iefc-baseline}, the best contrast achieved within 30 iterations is $7.0\times10^{-9}$. 

After obtaining the baseline iEFC result, we move on to perform iEFC while injecting WFE and closing the LLOWFSC loop to apply corrections. First, the LLOWFSC reference image is recorded and LLOWFSC is calibrated with the first 10 Zernike modes. Given the simulation results indicated that a modal basis that captures the higher order effects will be necessary to accurately compute the reference offsets, LLOWFSC is also calibrated with the same Hadamard basis as is used for iEFC. The calibration amplitude for all these modes was 3 nm. 

\begin{figure}[h!]
    \centering
    \includegraphics[scale=0.3]{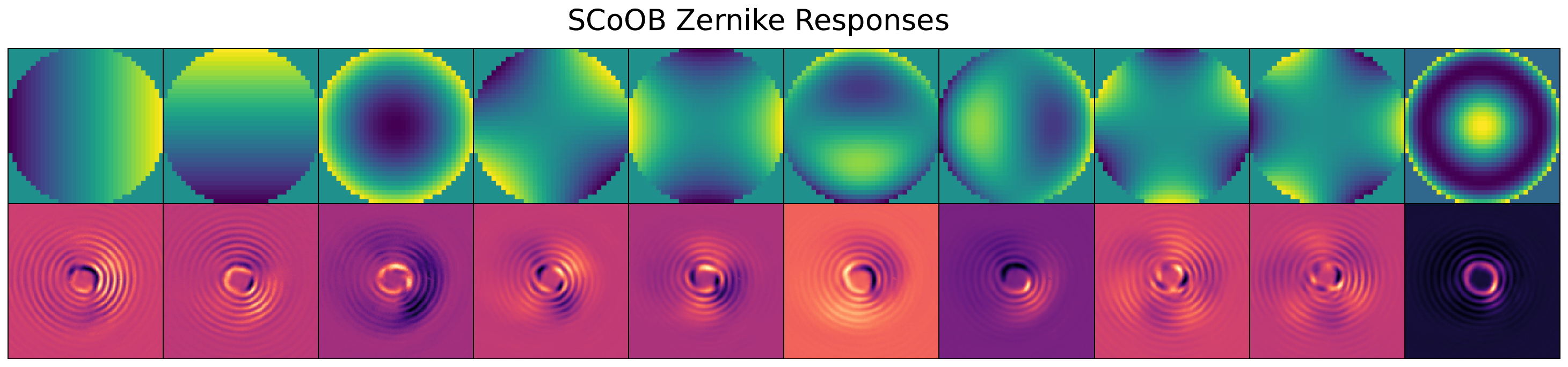}
    \includegraphics[scale=0.3]{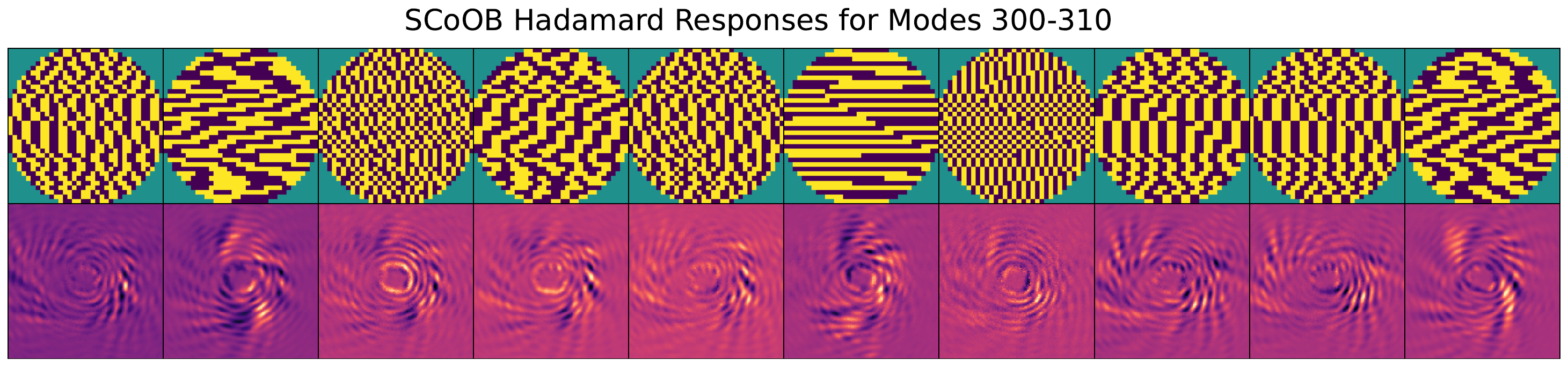}
    \caption{Similar to Figures \ref{fig:sim-llowfsc-responses} and \ref{fig:sim-hadamard-responses}, these are the LLOWFSC responses to the first 10 Zernike modes and a subset of 10 Hadamard modes with the SCoOB hardware.}
    \label{fig:scoob-llowfsc-calibs}
\end{figure}

Next, a time series of WFE is created by defining a PSD for both the tip/tilt modes and a PSD for the defocus - spherical modes similar to the method used for the simulation. However, the camera would only operate at 82 FPS with the ROI settings chosen for these experiments, so the knee frequency of tip/tilt is reduced to 0.1 Hz and the defocus - spherical modes are reduced to 0.01 Hz. This means the WFE injected for these experiments have less high temporal frequency components because CAMLO cannot operate fast enough to correct high frequencies. Additionally, the amplitude of the tip/tilt modes injected was set to 29.1 nm RMS because this is equivalent to 10 mas across a 2.4 m aperture. This value was chosen because the full pupil diameter for the STP ESC is 2.4 m. 

Due to computational limits of the desktop used to operate SCoOB, we only generate a time series for 15 min of data sampled at a rate of 500 Hz. This WFE is injected by using the kiloModulator application within SCoOB's software because it accurately injects the WFE at the desired rate of 500 Hz. Unfortunately, the application could not load data sets much larger than the 15 min time series, but this will be investigated and corrected for future experiments. After the final WFE sample is reached, the kiloModulator automatically restarts at the beginning of the time series again. This means the WFE injection can happen continuously for more than 15 min, but there is only 15 min of unique WFE being injected.  

\begin{figure}[h!]
    \centering
    \includegraphics[scale=0.4]{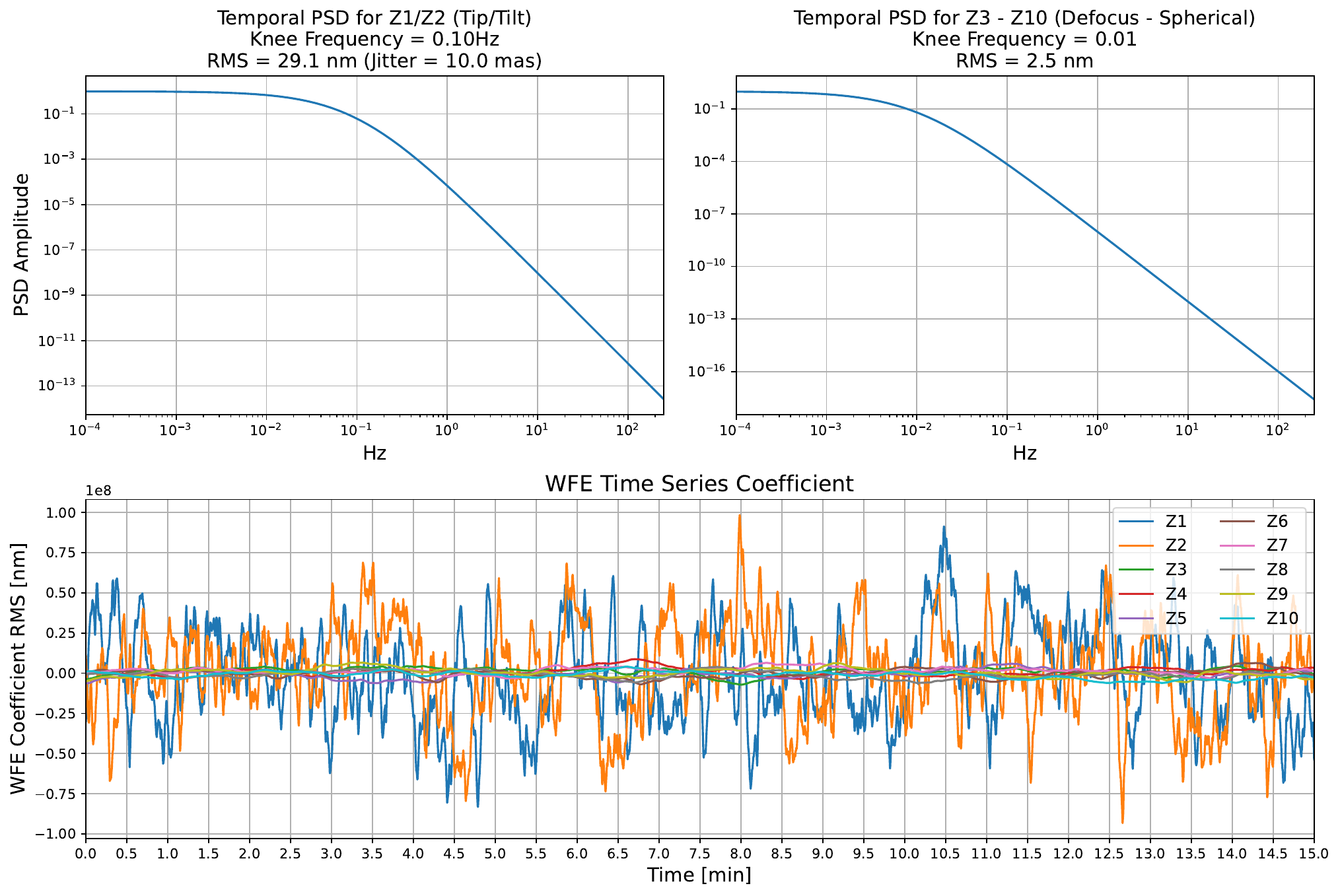}
    \caption{In the top row are the two separate PSDs for the tip/tilt modes and for the defocus - spherical modes used to create a random time series of WFE that is injected into SCoOB. The time series of coefficients (in terms of RMS) for each mode are shown in the bottom row.}
    \label{fig:scoob-wfe-time-series}
\end{figure}

\begin{figure}[h!]
    \centering
    \includegraphics[scale=0.525]{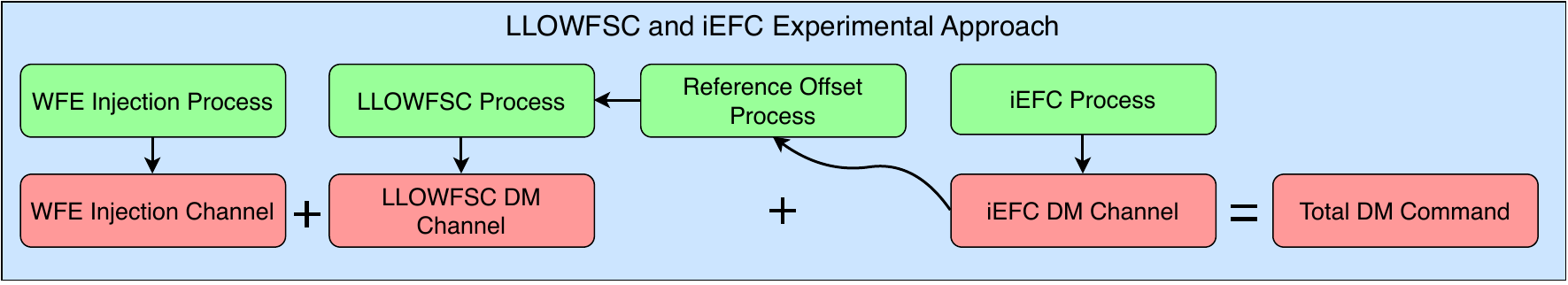}
    \caption{Illustration of the four separate processes defined to inject the WFE, operate LLOWFSC to compute and apply low order corrections, update the reference offset, and run iEFC. Specifically, the WFE injection applies the WFE to a designated DM channel while the LLOWFSC applies corrections to another channel. Similarly, iEFC applies dark hole commands to a channel monitored by the reference offsetting process such that a new offset is computed when the iEFC command is updated.}
    \label{fig:scoob-llowfsc-diagram}
\end{figure}

With the WFE time-series generated, the WFE injection and LLOWFSC processes are started. Simultaneously, we also begin the reference offsetting process which monitors the DM channel used for iEFC and computes the offset based on the current iEFC command. This includes computing the reference offset during iEFC calibration such that we test how well iEFC can be calibrated in the presence of dynamic WFE while LLOWFSC is on. Figure \ref{fig:scoob-llowfsc-diagram} illustrates these separate processes and how they communicate with the others. 

\begin{figure}[h!]
    \centering
    \includegraphics[scale=0.65]{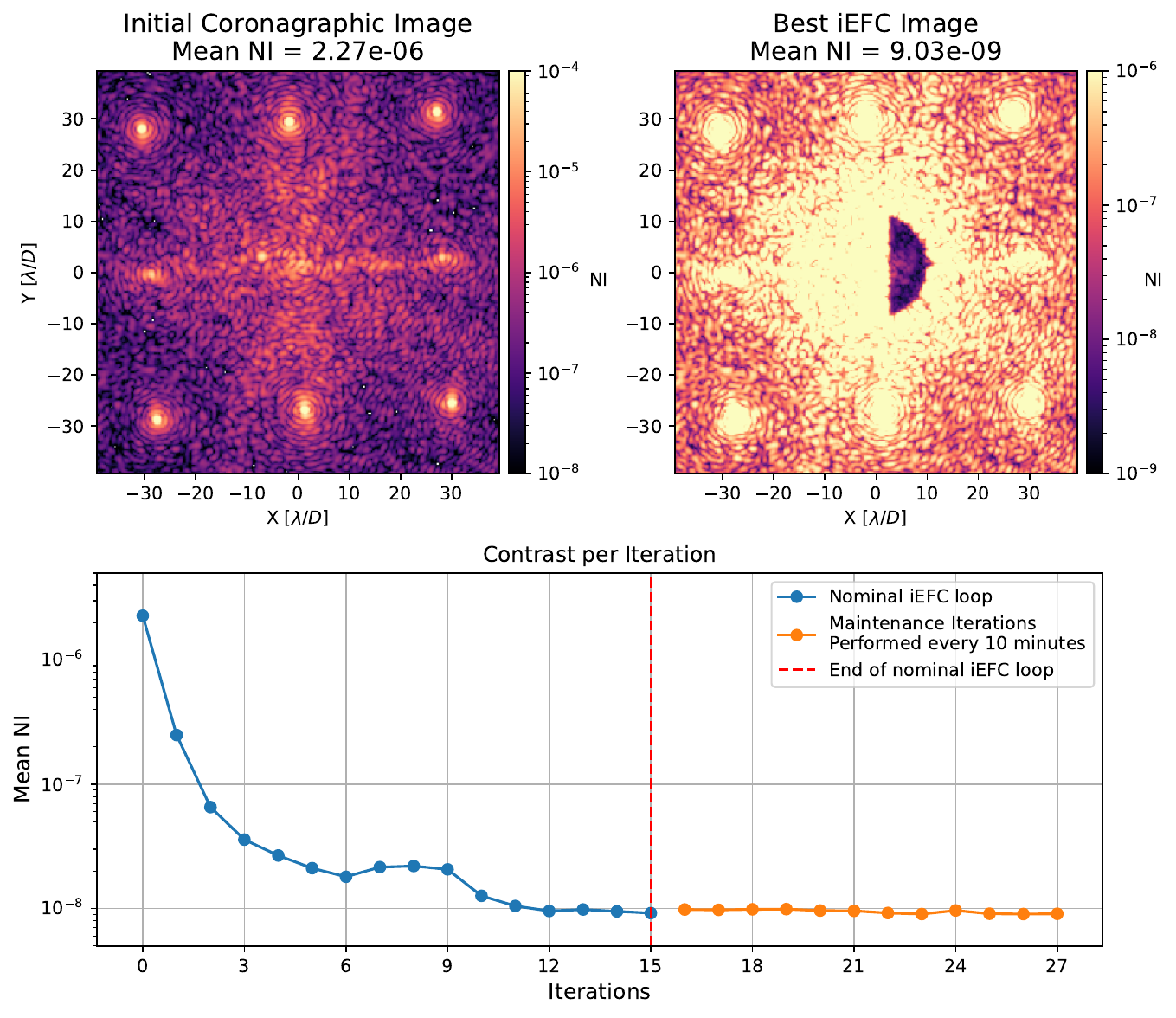}
    \caption{After closing the LLOWFSC control-loop and beginning the process that updates the reference offset, we calibrate and perform iEFC again to evaluate if LLOWFSC adequately corrects the dynamic WFE. In the top left is the initial coronagraphic image while the top right is the best contrast achieved. In the bottom are the contrast values per iEFC iteration, but after 15 iterations, the iEFC control-loop is set to run just once every 10 min while we evaluate the contrast maintained by LLOWFSC over each 10 min interval. At this stage, iEFC is acts as a maintenance algorithm, not as a dark hole creation algorithm.}
    \label{fig:scoob-llowfsc-iefc}
\end{figure}

After calibrating iEFC, we close this HOWFSC control-loop to generate a dark hole. Figure \ref{fig:scoob-llowfsc-iefc} presents the results and shows that the best contrast achieved is $9\times10^{-9}$. This indicates that the control-loop for LLOWFSC and the reference offset process are functioning as expected, but the corrections are not adequate enough to achieve the same performance as the baseline iEFC case. 

\begin{figure}[h!]
    \centering
    \includegraphics[scale=0.325]{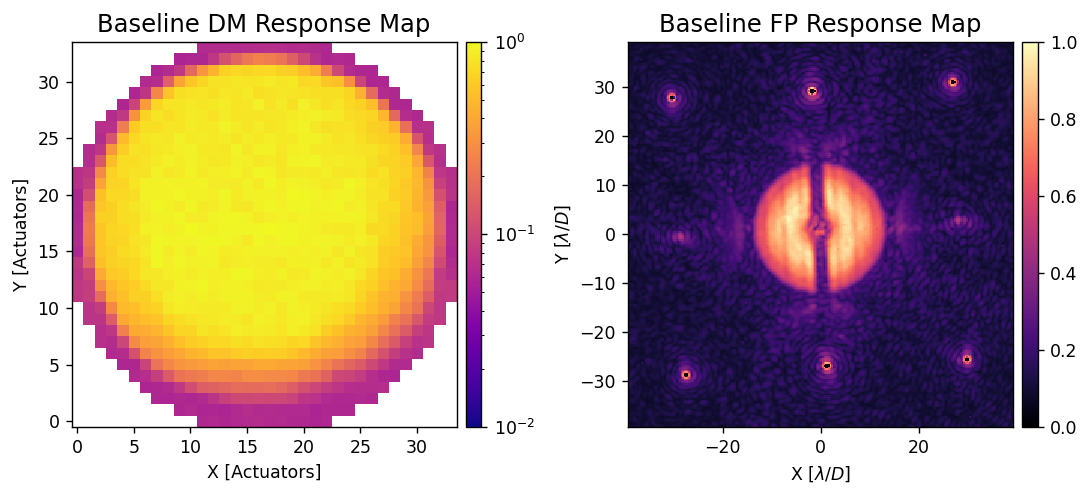}
    \includegraphics[scale=0.325]{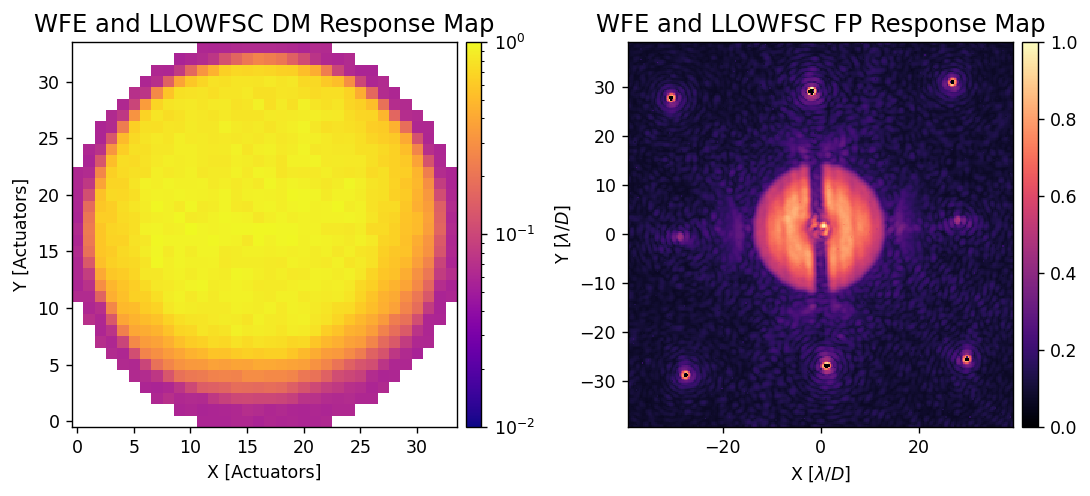}
    \caption{The iEFC response maps from the the baseline calibration (top) and the calibration measured with WFE injection and LLOWFSC turned on (bottom). While the response maps on the DM actuators appear similar, the focal plane response maps appear different because there are brighter components appearing similar to speckles inside the IWA. This is an indication the calibration accuracy is degraded with the dynamic WFE injection and can limit contrast performance. }
    \label{fig:scoob-iefc-response-maps}
\end{figure}

This is likely due to a degraded iEFC calibration as Figure \ref{fig:scoob-iefc-response-maps} shows the difference in the focal plane response maps from the baseline iEFC calibration and the calibration while injecting WFE. While the response maps on the DM actuators appear similar, the focal plane response maps appear different because there are brighter components appearing similar to speckles inside the IWA. This is an indication the calibration accuracy is degraded with the dynamic WFE injection, so it prevents the dark hole from reaching the same contrast as the baseline case. This could be caused by inaccurate LLOWFSC corrections due to measurement noise, time lag, nonlinearity of the responses, or coupling between low-order modes. Note that the DM response maps are expected to be biased toward one side of the DM because we only create a half dark hole, but the focal plane response maps are mostly symmetric because the Fourier probes have a symmetric response. 

\begin{figure}[h!]
    \centering
    \includegraphics[scale=0.6]{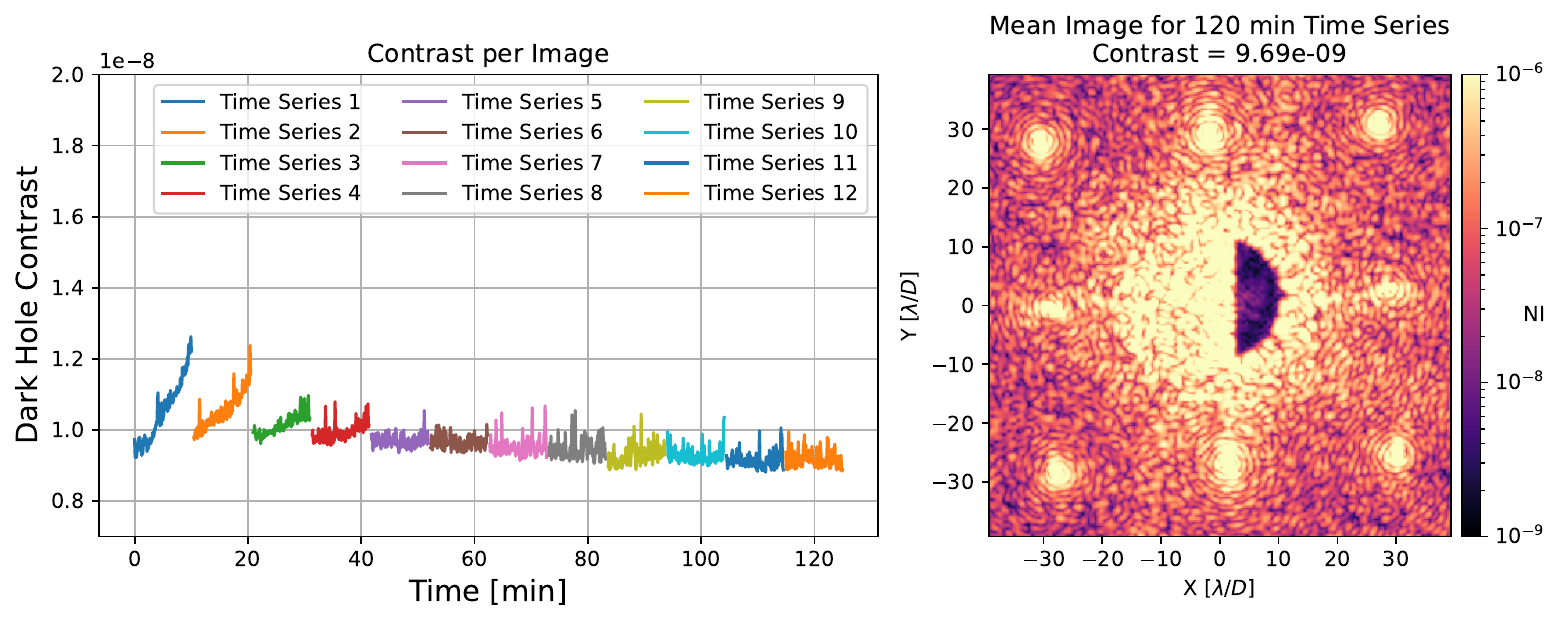}
    \caption{To the left is a plot of the contrast versus time for each of the 10 min intervals previously indicated in Figure \ref{fig:scoob-llowfsc-iefc}. There is a break between each 10 min interval because iEFC is used to maintain the dark hole in the case of drifts. It is suspected that these drifts were dominant during the first few time series because the data was collected on a Friday afternoon and most movement near the laboratory space occurs during the first few intervals. To the right is average dark hole across the entire 120 min period demonstrating that a contrast below $10^{-8}$ was maintained.}
    \label{fig:scoob-llowfsc-time-series}
\end{figure}

Also shown by Figure \ref{fig:scoob-llowfsc-iefc} is that after the first 15 iterations of iEFC, the loop is set to perform one iteration every 10 min while we evaluate the contrast maintained by LLOWFSC during those 10 min periods. Figure \ref{fig:scoob-llowfsc-time-series} presents those contrast results where over the course of the complete 120 min interval, the average contrast maintained is $9.7\times10^{-9}$. During this time period, the LLOWFSC process was operated with no leakage, a gain of 0.5 on the tip/tilt modes and 0.1 on the defocus-spherical modes. Notably, the first few maintenance periods have the highest degree of contrast degradation. The exact cause of this degradation is unknown, but the current theory is that it was caused by higher order drifts due to instability of the laboratory environment which settled after about 40 min. The data was collected on a Friday afternoon when there is the most movement of people leaving the offices near the laboratory, so it is possible that the high order drifts were due to outside movement. However, the iEFC iterations recover the contrast to below $10^{-8}$ after each of the periods with significant degradation. 

\begin{figure}[H]
    \centering
    \includegraphics[scale=0.6]{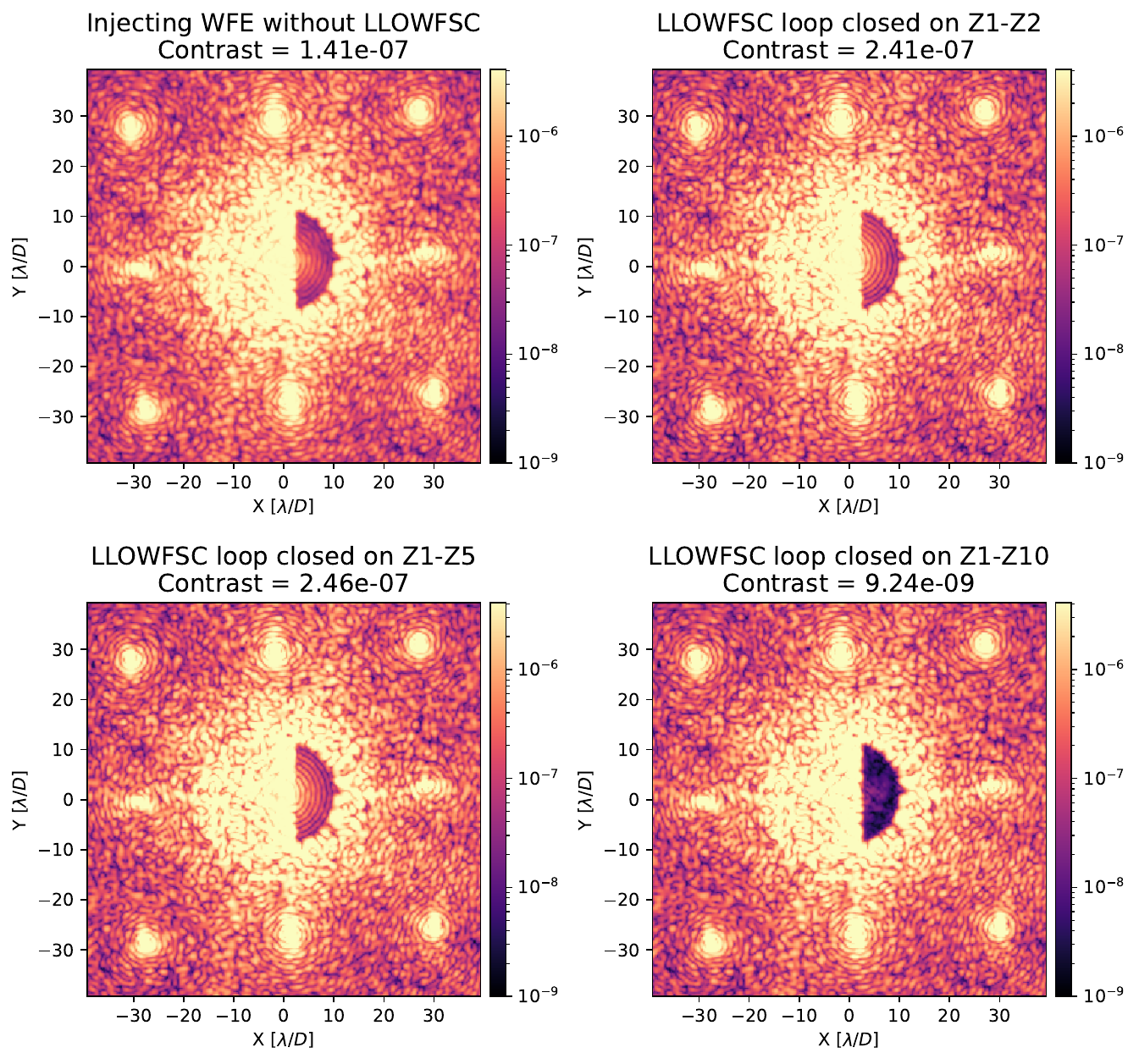}
    \caption{These images demonstrate the contrast degradation from the WFE injection. In the top left is the mean dark hole image with the WFE injection on, but the LLOWFSC correction off. In the top right is the dark hole image with the LLOWFSC loop closed only on the tip/tilt modes, then closed on the first 5 Zernike modes in the bottom left, and finally closed on all 10 Zernike modes in the bottom right. Once the loop is closed on all modes, the contrast in the dark hole is quickly recovered.}
    \label{fig:scoob-llowfsc-open-vs-closed}
\end{figure}

In Figure \ref{fig:scoob-llowfsc-open-vs-closed}, we can also see the impact of the low-order Zernike aberrations that are being injected. Without closing the loop on all the Zernike modes, the contrast in the dark hole is above $10^{-7}$. The leakage from these low-order modes appear similar to Airy rings in the final dark hole, but once the loop is closed on all modes, these leakage terms are no longer visible above the signal of the speckles. 

\begin{figure}[H]
    \centering
    \includegraphics[scale=0.6]{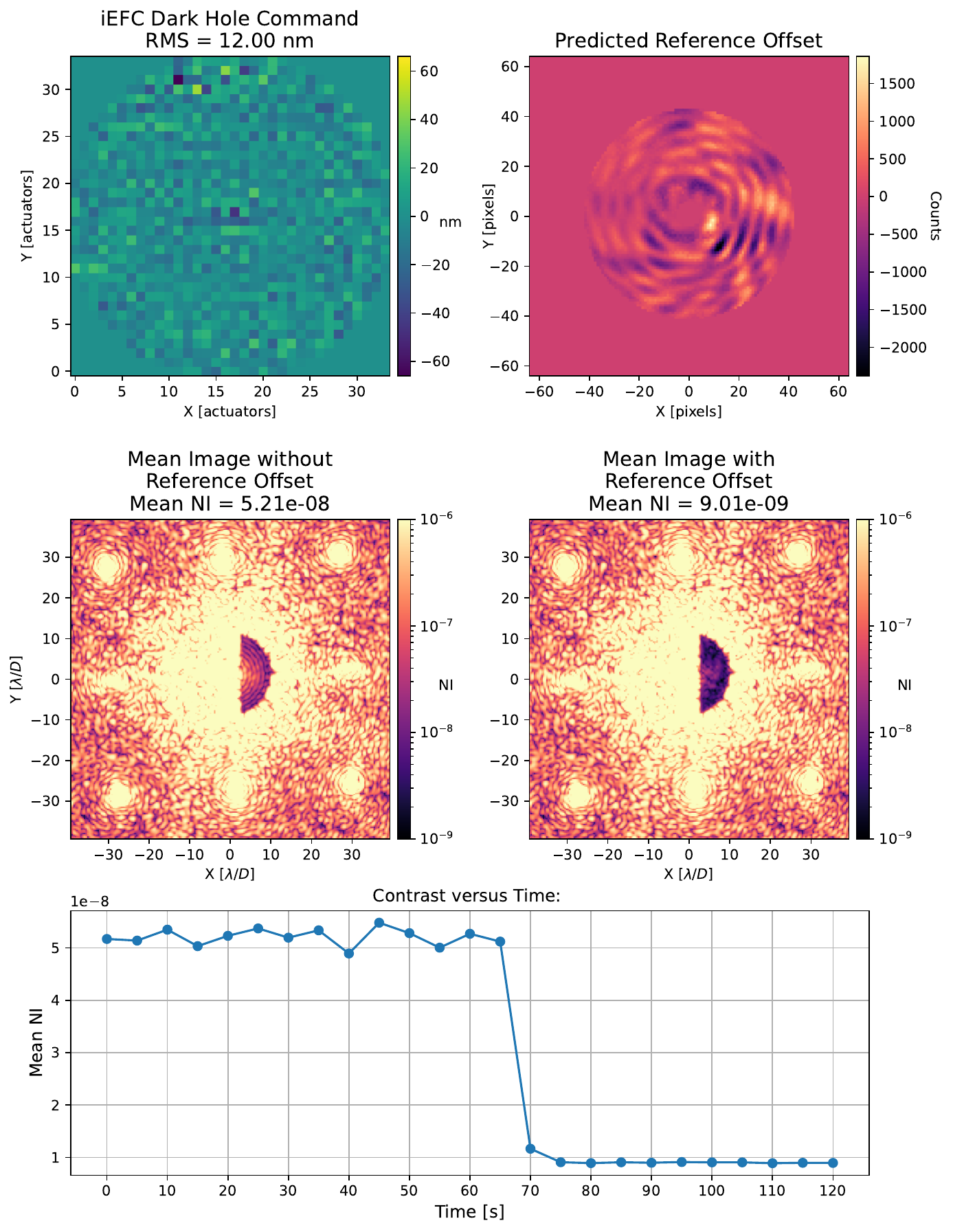}
    \caption{In the top left is the final dark hole command from iEFC followed by the reference offset computed in the top right. In the middle left is average dark hole image with the reference offset turned off while in the middle right is the dark hole after turning the reference offset on. The bottom shows the contrast versus time where the reference offset is turned back on after about the 65 s mark to demonstrate the importance of this step for combining LLOWFSC and HOWFSC.}
    \label{fig:scoob-llowfsc-ref-offsets}
\end{figure}

Lastly, Figure \ref{fig:scoob-llowfsc-ref-offsets} evaluates the impact of the reference offset. If the offsetting process is deactivated, a similar low-order leakage term is produced in the dark hole due to LLOWFSC implicitly removing the low-order components of the iEFC command. Once the offsetting process is reactivated, the dark hole contrast is quickly recovered. The reference offset computed throughout these closed-loop tests on SCoOB utilized the complete Hadamard calibration of the LLOWFSC signal. No tests were done with a subset of Hadamard SVD modes like those shown with the simulations due to time constraints with SCoOB, so this is left as future work. 

\section{Conclusions}
\label{sec:llowfsc-conclusions}

This work has outlined how a LLOWFSC and HOWFSC method can be combined to achieve and maintain high-contrasts with a vortex coronagraph. This will be critical for VVC concepts that seek to achieve these high-contrasts in order to image exoplanets since HOWFSC methods require highly stable wavefronts to achieve optimal performance. The experiments with SCoOB have demonstrated the maintenance of sub-$10^{-8}$ contrast at 633 nm with a monochromatic source. Critically, we show that the data driven HOWFSC method of iEFC can be calibrated and closed in the presence of dynamic WFE if LLOWFSC is used to maintain stability.  

Future work will include testing with a broadband source to simulate operation on target stars with flight-like flux levels. Additionally, we will be upgrading the camera hardware to the flight-like cameras and testing the LLOWFSC control-loop operating at higher frequencies. In turn, this will allows us to expand the LLOWFSC tests for different WFE dynamics including higher knee frequencies for jitter and other low-order Zernikes. 

\acknowledgments % equivalent to \section*{ACKNOWLEDGMENTS}   
 
This unnumbered section is used to identify those who have aided the authors in understanding or accomplishing the work presented and to acknowledge sources of funding.  

% References
\bibliography{refs_scoob_llowfsc} % bibliography data in report.bib
\bibliographystyle{spiebib} % makes bibtex use spiebib.bst

\end{document}